\documentclass[11pt]{article}
\usepackage[american]{babel}
\usepackage{geometry}
\usepackage{setspace} 
\usepackage{lscape} 

\usepackage{graphics}

\usepackage{apacite}
\usepackage{makerobust}
\DeclareRobustCommand{\citet}[2][]{\citeA[#1]{#2}}
\MakeRobustCommand{\citeauthor}

\usepackage{mathtools}
\usepackage{bm}
\usepackage{amsmath}
\usepackage{amssymb}
\usepackage{blkarray}

\usepackage{booktabs}
\usepackage{multirow}
\usepackage{array}
\newcommand{\rowgroup}[1]{\hspace{-1em}#1}
\usepackage[flushleft]{threeparttable}

\usepackage[toc,page]{appendix}

\usepackage{xurl} 

\usepackage[normalem]{ulem} 
\usepackage{xcolor} 

\usepackage{authblk}

\title{A Censored Mixture Model for Modeling Risk Taking}
\author[1]{Nienke F.S. Dijkstra}
\author[2]{Henning Tiemeier}
\author[3]{Bernd C. Figner}
\author[1]{Patrick J.F. Groenen}
\affil[1]{Erasmus University Rotterdam}
\affil[2]{Harvard T.H. Chan School of Public Health}
\affil[3]{Radboud University}
\date{\today}

\begin{document}
	\maketitle
	\noindent\textbf{Corresponding author:} Nienke Dijkstra, n.f.s.dijkstra@ese.eur.nl
	
	\begin{abstract}
		Risk behavior can have substantial consequences for health, well-being, and functioning. Previous studies have shown an association between real-world risk behavior and risk behavior on experimental tasks, such as the Columbia Card Task, but their modeling  is challenging for several reasons. First, many of the experimental risk tasks may end prematurely leading to censored observations. Second, certain outcome values can be more attractive than others. Third, a priori unknown groups of participants can react differently to certain risk-levels. Here, we propose the Censored Mixture Model (CMM), which models risk taking while handling censoring, experimental conditions, and attractiveness to certain outcomes.
	\end{abstract}
	
	
	\section{Introduction}
Taking a particular risk can have substantial consequences on health, well-being, and functioning. Consequently, risk taking is examined in multiple scientific fields, such as psychology, criminology, and economics, and is measured by surveys or using experimental tasks. Although risk behavior in different experimental tasks do not highly correlate \cite{pedroni2017risk}, studies have shown a moderate, but meaningful association between risk behavior measured in various experimental tasks and real-world risk taking. For example, \citet{lejuez2003balloon} and \citet{pripfl2013effects} show that smokers take significantly more risk on respectively the Balloon Analogue Risk Task (BART) and the Columbia Card Task (CCT) than non-smokers. Likewise, \citet{collins1987psychosocial} show the relationship between risk taking/rebelliousness, assessed with a survey, and smoking at an older age.

There are four types of experiments commonly used to measure risk behavior. The first is based on lotteries, where an explicit description of the outcome and probabilities is given. Typically, participants have to state their preference between, for example, option A: 50\% chance of winning 10 euro, and option B: 30\% chance of winning 30 euro. Typically, in these tasks it is straightforward to decompose the underlying constructs of risk taking. However, they are often criticized for being too artificial and lacking external validity.

An example of the second type of experiments is the Iowa Gambling Task \cite{bechara1994insensitivity}, where participants can win or lose money by picking (many) cards from four decks, each card having a win and loss value. The expected value and probability distribution of the values of the cards in the four decks are unknown to participants at the start, but can be learned during the task. This task has shown to successfully predict real-world risk taking behavior. However, it is virtually impossible to decompose the underlying constructs of participants' risk taking behavior, such as risk preferences \cite{schonberg2011mind}. Risk preferences are confounded with the learning curve, because participants have to unravel the expected return and the probability distribution of the decks while playing the game. In addition, it is difficult to distinguish whether the behavior is driven by risk attitude or sensitivity to reward or punishment.
	
The third type of task paradigm is based on gambling and includes, among others, the Cambridge Gambling Task \cite{rogers1999choosing} and the Game of Dice Task \cite{brand2005decision}, where participants have to bet on the color of randomly drawn cards or on the outcome of a roll of a dice, respectively. The probability of the possible outcomes is known, so there is no learning effect present. However, these tasks have the disadvantage that they do not allow to disentangle the effects of risk and of attractiveness of a higher expected pay-off value.

Last, in sequential risk tasks, such as the Balloon Analogue Risk Task (BART) \cite{lejuez2002evaluation} and the Columbia Card Task (CCT) \cite{figner2009affective}, participants are asked to repeat a certain action (e.g., inflating a balloon or turning over cards). The risk increases the longer a participant continues. Although the BART and CCT do not suffer from the issues described above, they have their own challenges, which makes modeling risk taking complex.
	
First, the analyses are often based on the assumption of a smooth (normal) distribution of the residuals. However, certain outcomes are more attractive to participants than others. For example, in some sequential risk tasks participants have to select a number of repetitions of a certain action, this number indicates the level of risk someone is prepared to take (e.g., the number of pumps in the BART or the number of cards turned over in the CCT). It is well known that even numbers and multiples of five are more often selected than odd numbers \cite{baird1970relative}. This pattern leads to inflated values in the outcome distribution. Similarly, within surveys some outcomes tend to be more attractive than others. Imagine a longitudinal study measuring drug usage. Asking people how often they use drugs, typically, also leads to even numbers and multiples of five and ten \cite{klesges1995self}.

The second challenge of sequential risk tasks concerns censored observations. In the imaginary longitudinal drug study, participants can be easily lost, leading to incomplete information and censored observations. Moreover, most sequential risk tasks by definition may randomly end prematurely, such as for the BART and CCT. Typically, the researcher is interested in the level of risk a participant is willing to take and the censoring obscures this. 
One solution for dealing with censored observations was proposed by \citet{lejuez2002evaluation}. They suggest to use the adjusted score in the BART (average inflations over the unpopped balloons). However, \citet{pleskac2008development} have shown that this score is biased and propose the automatic BART, where participants have to choose a number of inflations before the trial starts and censoring is no issue. This new version of the BART measures risk taking in a more planned situation, because participants do not feel the excitement of a successful inflation. \citet{figner2009affective} argue that people behave differently in a deliberative and affective decision making situation. The authors found that adolescents showed a greater increase in risk taking in the hot-affective compared to the cold-deliberative CCT, compared to the adults who showed more similar risk-taking levels across the hot and cold task versions.



\citet{figner2009affective} developed the Columbia Card Task to investigate the difference between deliberative and affective decision making. The CCT is a computer-based card game and participants can win or lose money by turning over cards. A major advantage of the CCT, over other dynamic risk tasks, is that the CCT orthogonally varies the risk-relevant factors gain amount, loss amount, and loss probability. Such an unconfounded design allows for the decomposition of risk taking into underlying mechanisms such as sensitivity to gains, losses, and probabilities \cite{figner2009affective, penolazzi2012impulsivity}.

The challenges of modeling risk taking also arise in the analysis of the CCT. The CCT has 32 cards divided in win cards and loss cards. By turning over a win card the participant collects points. However, if a loss card is encountered, the game round ends prematurely and a specified number of points is subtracted from the score of the respective game round. Turning over a loss card means that the researcher does not know the intended number of cards and this constitutes a censored observation. Note that the current study focuses on affective decision making and thus concerns only the hot CCT. Furthermore, the CCT may also suffer from inflated values in the distribution, because some participants create geometric patterns for turning over cards and thus find certain outcome patterns such as a single row or column of cards turned over more attractive than others.

So far, none of the existing studies have provided a statistical model that addresses all the issues introduced above \cite<though recent work addresses the censoring in the CCT,>{weller2019information}. Here, we propose a Censored Mixture Model (CMM). The censored observations are included in the model by using the information that the participant intended to take more risk than the observed level. The attractiveness of certain patterns in outcome values is covered by assigning extra probability mass to the inflated values in the distribution. Furthermore, the unobserved individual tendency for risk taking can be taken into account with finite mixtures. In addition, we choose a link function such that the regression coefficients have a linear interpretation on the interval $[0, \inf \rangle$.

The remainder of this paper is structured as follows. It starts with a detailed explanation of the CCT and its challenges when modeling risk taking. Next, the data is discussed by means of the data collection process, cleaning procedure, and their characteristics. Subsequently, the structure of the model is discussed extensively. Last, we present the results and we will discuss the limitations.

\section{Columbia Card Task}
\begin{figure}[h]
	\centering
	\includegraphics[width=0.8\linewidth]{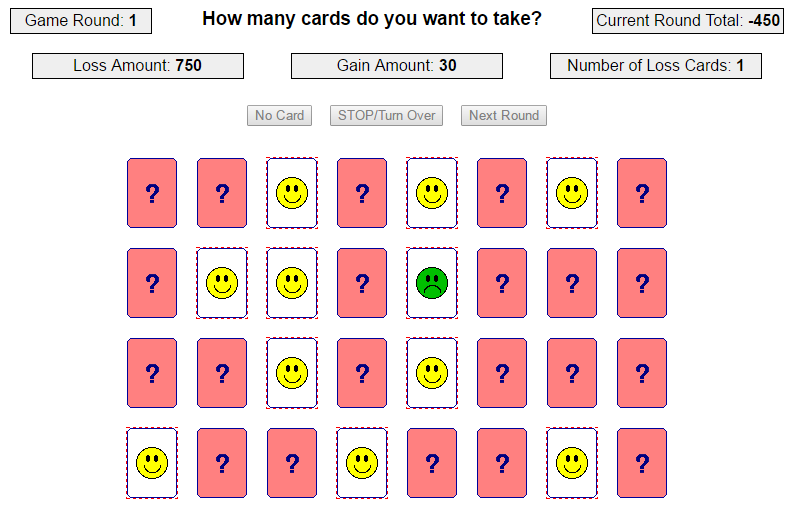}
	\caption{A screenshot of the first game round in the Columbia Card Task with the game settings: gain amount equal to thirty, loss amount equal to 750, and number of loss cards equal to one. In this game round, the participant first turned over ten win cards (happy faces). The eleventh card was a loss card (sad face), resulting in a total score in the current game round of $10 \times 30 - 750 = -450$.}
	\label{fig:CCT}
\end{figure}

The Columbia Card Task is shown in Figure \ref{fig:CCT}. There are 32 cards divided in win cards (happy faces) and loss cards (unhappy faces). At the beginning of a game round all cards are face down and participants are asked to turn over cards. By turning over a win card the participant earns points and by turning over a loss card they lose points and the current game round ends. At every step the participant has the choice between turning over another card and pressing the stop button to voluntarily stop this game round. It is also possible to stop immediately without turning over any card. After a game has ended the earned points are summed and the potential loss amount is subtracted.

The gain amount (points earned by turning over a win card), loss amount (points lost by turning over a loss card), and number of loss cards vary per round: the gain amount is either ten or thirty, the loss amount is either 250 or 750, and there are either 1 or 3 loss cards in the game \cite{figner2011takes}. These three experimental conditions are displayed at the top of the screen and are known to the participant. Note that, in contrast to \citet{figner2009affective}, the loss cards are randomly distributed over the 32 cards. These three parameters lead to eight different game settings and within a block of eight trials the sequence of the game settings is random. Every participant plays at least two blocks of eight trials\footnote{At the beginning of the data collection we decided to shorten the test. Instead of three blocks with in total 24 trials, two blocks with in total 16 trials were played. There are 388 children, who played 24 trials. For the analysis only the first 16 trials of these children are used.}. In other words, every game setting is played at least twice. Because of the different game settings, the CCT measures next to risk taking also the complexity of information use and the sensitivity to reward, punishment, and probability. With the three parameters (gain amount, loss amount, and number of loss cards) is it possible to assess which of these three parameters affects participants' choices.

The indicator for risk taking is the number of cards a participant intends to turn over. However, if a participant faces a loss card the game ends prematurely, the trial is censored and it is unknown how many cards a participant intended to turn over. This should be considered in the analysis. \citet{figner2009affective} manipulate the game such that in most trials the loss card is the last possible card to turn over and only analyze the uncensored trials. However, for this manipulation not to be discovered by participants, \citet{figner2009affective} included extra trials where the loss card appeared at an early stage of the trial. This approach has several drawbacks. Besides the serious problem that such a setup uses deception, letting participants play extra rounds has the important disadvantage of being time consuming and hence more expensive. In addition, we show that the result in the previous trial effects the behavior in the current trial. Not correcting for the negative experience of facing a loss card, could affect the results.

\begin{figure}[h]
	\centering
	\includegraphics[width=1\linewidth]{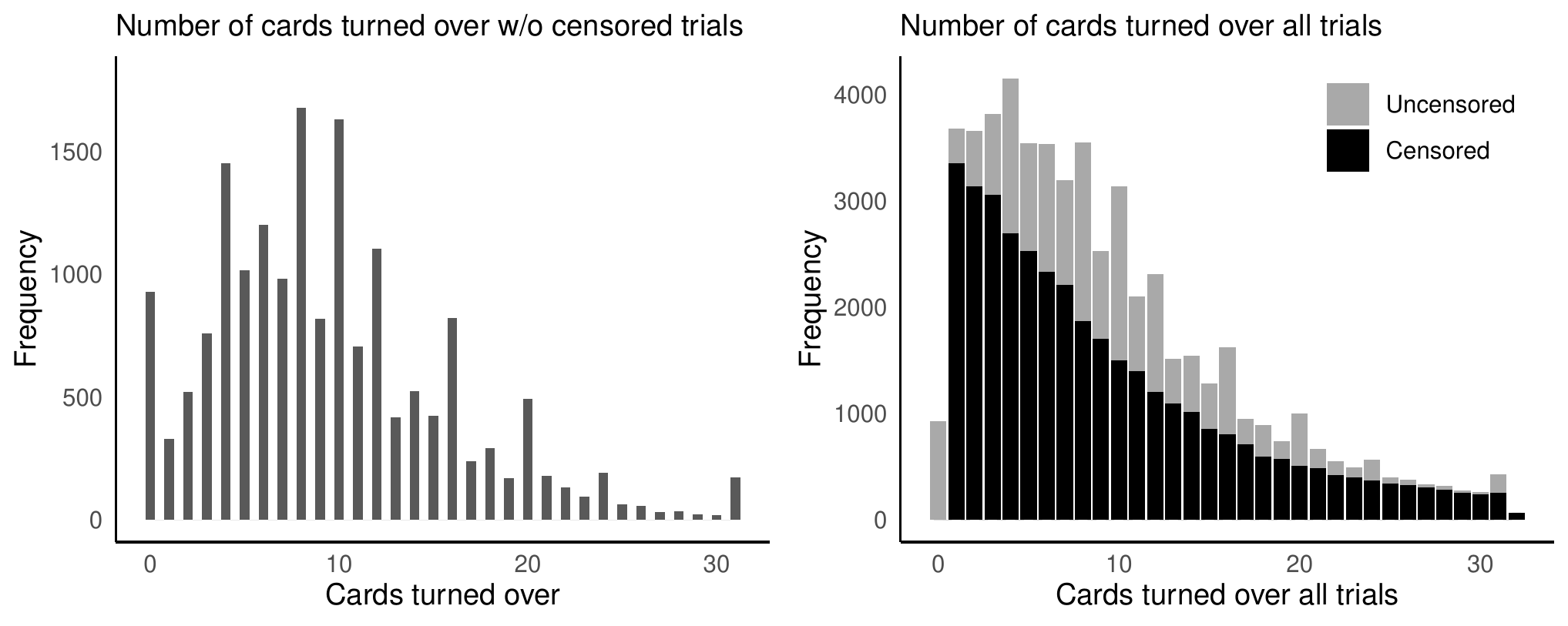}
	\caption{Distribution of the number of cards turned over}
	\label{fig:numCards}
\end{figure}

Another issue that should be accounted for is the attractiveness of certain outcome values. Figure \ref{fig:numCards} shows the distribution of the outcome, in this case, the number of cards turned over. The left graph only includes the uncensored trials and shows peaks at certain number of cards. The right panel suggests that the peaks are independent of the probability of being censored, because the censored trials (lower bars) do not show any irregular or unexpected values. Three categories of peaks can be distinguished. The first category is the excess of zeros. This inflation is probably caused by children who are very much risk averse and prefer not to play the game. The second category of excesses occurs with participants who are very risk seeking: if you managed to turn over 30 cards without facing a loss card, then why not as well try the 31\textsuperscript{st} card? The third category includes the peaks at four, eight, ten, twelve, sixteen, twenty, and twenty-four. Although ten is not a multiple of four, it seems to be an attractive number similar to the multiples of four, hence it is included in this set. Recall the layout of the CCT from Figure \ref{fig:CCT}, creating a geometric pattern, such as complete rows or columns, corresponds to turning over a number of cards equal to a multiple of four.

\section{Data}
\label{sect:data}
What sets this research apart from previous studies, besides addressing all issues involved with modeling risk taking, is the large number of participants. The large sample size allows us to build a more flexible model that handles censoring, categorical background variables, such as individual characteristics, experimental conditions, and attractiveness to certain patterns and outcome values.

The current study is embedded in the Generation R Study, a large population based multi-ethnic cohort study \cite{kooijman2016generation}. The Generation R Study was designed to analyze early environmental and genetic determinants of growth, development, and health from foetal life until young adulthood. The data collection is intense, includes multiple surveys with biological and observational assessments. The CCT is one of the observational assessments that was conducted on nine-year-old children (age 9.8 $\pm 0.26$). The cohort includes almost ten thousand children at birth, of which 4538 children participated in the CCT. The data set is partitioned in a training set of 3404 children and an (prior to analysis) unseen test set of 1134 children. The Generation R study has an open policy in regard to collaboration with other research groups (\url{http://www.generationr.nl/researchers/collaboration.html}). Requests for data access and collaboration are discussed in the Generation R Study Management Team.

The CCT was conducted as part of a series of assessments taking approximately three hours. At the beginning of the CCT, children were told that they would be rewarded with money based on their performance on the CCT. After all trials were played, three trials were randomly selected and were paid out in real money. The children had a start value of 200 cents (i.e., 2 euro) and the total points of the selected trials were added or deducted from this start value. Children could receive money, but did not have to pay any net losses.

The prevalence of censoring (i.e., the number of observation with incomplete data) in this data set is $68\%$. Therefore, treating the censored observations as uncensored would lead to severe biases in the results. Available background variables include children's age and IQ (102 $\pm 14.7$), measured with the SON-R 2.5-7 at the age of six. Furthermore, information about the mother is available in ethnicity (Dutch = 59.8\%, Dutch Antilles = 2.1\%, African and Moroccan $<5\%$, Asian (non Western) and Turkish $<6\%$, Surinamese = 7.1\%, and other Western = 10,1\%) and education (low = 6.7\%, middle = 42.2\%, high = 51.2\%). The last background variable is the household income per month in euros ($<2000 = 20.5\%$, $2000-4000 = 43.8\%$, $>4000 = 35.7\%$). Missing values in the background variables are imputed with single Predictive Mean Matching (PMM) using age, gender, weight at birth, and IQ of the child, and the age at delivery, ethnicity, and education of the mother, and household income as predictors, and using the \texttt{mice} package in \texttt{R}.

The segments obtained with the CMM will be interpreted using the child behavior checklist (CBCL). This survey assesses child emotional and behavioral problems as perceived by the mother. The CBCL has been completed at the same time as the CCT, at age nine. For some children, the CBCL scores at age nine were missing but available at age six. For these children, their scores at age nine are imputed by single Predictive Mean Matching (PMM) using the score at age six and the covariates in the model. The 223 children that have scores neither at age six nor at age nine are excluded from this analysis. 

\section{Methods}
The following section is concerned with the methods and techniques applied in this study. First, the structure of the model is discussed, extensively. Several challenges emerging in modeling risk taking by the CCT are discussed when construing the likelihood function. Additionally, we discuss a mean zero restriction for the weights estimated for a categorical variable instead of the usual reference category.

\subsection{The Censored Mixture Model (CMM)}
To model the CCT data, three challenges need to be taken care of: the censoring of the data, the attraction of particular outcomes, such as presented in Figure \ref{fig:numCards}, and unobserved heterogeneity across individuals. Several challenges of modeling risk taking are addressed by the CMM. The possible censoring is accommodated by the cumulative distribution function, which is added to the likelihood function. Extra probability mass is assigned to the inflated values in the outcome distribution. Last, the finite mixtures account for the unobserved individual characteristics. Apart from this, we follow a generalized linear model approach, that is, we will assume that there is a linear combination of covariates that provides, after transformation by a link function, the mean of a distribution for every observed number of cards. We argue that the negative binomial distribution is appropriate and provide a link function that is close to linear for ease of interpretation. Below, a step wise explanation is given how these potential problems are solved by the CMM.

The observed variable to be modeled is the number of cards turned over $y_{it}$ by individual $i$ in trial $t$. Furthermore, we observe whether a trial is censored at card $k$, $c_{itk} = 1$, or not, $c_{itk} = 0$. However, we are interested in the latent random variable $Z_{it}$, indicating the number of cards someone intends to turn over. This variable is assumed to follow a known distribution (here we propose to use the negative binomial distribution). Now, the probability of the observed number of cards turned over can be expressed in terms of the latent random variable $Z_{it}$
\begin{equation*}
\label{eq:prob}
\Pr(Y_{it} = k \wedge C_{itk} = c_{itk}) = \Pr(Y_{it} = k \wedge C_{itk} = c_{itk} \mid Z_{it} = \ell) \Pr(Z_{it} = \ell),
\end{equation*}
where $C_{itk}$ indicates censoring at trial $t$ for individual $i$ at card $k$.


More insight on the conditional probability $\Pr(Y_{it} = k \wedge C_{itk} = c_{itk} \mid Z_{it} = \ell) $ can be obtained by considering all possible outcomes of the game, that is, for all combinations of number of observed cards $y_{it}$, being censored or not ($c_{itk} = 1$ or 0), and the number of cards intended to turn over $z_{it}$. Table \ref{tab:prob} provides these probabilities where the notation $p_k = \Pr(C_{itk} = 1 \mid C_{it\ell} = 0 \ \forall \ \ell < k)$ is used. 	Note that these probabilities are purely based on the game settings. Due to symmetry properties, this table can be summarized by
\begin{eqnarray*}
	\Pr(Y_{it} = k \wedge C_{itk} = c_{itk} \mid Z_{it} = \ell)
	&=&
	\left\{ \begin{aligned}
		& \Omega_{k\ell, c_{itk}} = \Omega_{kk, c_{itk}}  &&\forall \ell \geq k \text{ if } c_{itk} = 1,\\
		& \Omega_{k\ell, c_{itk}} = 0 \ \ \ \ &&\forall \ell < k\text{ if } c_{itk} = 1,\\
		& \Omega_{k\ell, c_{itk}} = 0 \ \ \ \ &&\forall \ell \neq k \text{ if } c_{itk} = 0\\
	\end{aligned} \right.
\end{eqnarray*}
where $\Omega_{k\ell, c_{itk}}$ corresponds to the fixed probabilities given in Table \ref{tab:prob}. Subsequently, the likelihood contribution for person $i$ at trial $t$ can be written as
\begin{equation*}
\begin{aligned}
L_{it} &= \sum_{\ell = 0}^{32} \Pr(Y_{it} = k \wedge C_{itk} = c_{itk} \mid Z_{it} = \ell) \Pr(Z_{it}=\ell)\\
&= \left\{ \begin{aligned}
&\Omega_{k\ell, c_{itk}} \Pr(Z_{it} = \ell) = \Omega_{k\ell, c_{itk}} \Theta_{\ell, c_{itk}} && \text{if } c_{itk} = 0 \\
&\sum_{m = k}^{32} \Omega_{km, c_{itk}} \Pr(Z_{it} = m) = \Omega_{k\ell, c_{itk}} \sum_{m = k}^{32} \Pr(Z_{it} = m) = \Omega_{k\ell, c_{itk}} \Theta_{\ell, c_{itk}} && \text{if } c_{itk} = 1.
\end{aligned}\right.
\end{aligned}
\end{equation*}
Thus, for $c_{itk} = 1$, $\Theta_{\ell, c_{itk}}$ equals one minus the cumulative distribution function (cdf), that is, $\Theta_{\ell, c_{itk}} = 1 - \sum_{\ell =0 }^{k-1} \Pr(Z_{it} = \ell)  $ and for $c_{itk} = 0$ we have $\Theta_{\ell, c_{itk}} = \Pr(Z_{it} = \ell)$. Multiplying over the trials, the likelihood contribution of person $i$ can be written as
\begin{equation*}
\begin{aligned}
L_i &= \prod_{t = 1}^{T} \Omega_{y_{it}z_{it}, c_{itz_{it}}} \Theta_{z_{it}, c_{ity_{it}}}.
\end{aligned}
\end{equation*}

\begin{table}
	\footnotesize
	\label{tab:prob}
	\caption{The probability for each possible combination of the  observed number of cards $k$, the intended number of cards $\ell$, and  being censored at card $k$ ($c_{itk}$), that is, $\Pr(Y_{it}~=~k \wedge C_{itk}~=~c_{itk} \mid Z_{it}~=~ \ell) = \Omega_{k\ell, c_{itk}}$.}
	\begin{center}
		\begin{tabular}{@{}cl|llllllll@{}}
			\hline
			&   & \multicolumn{6}{c}{$z_{it} =\ell $}                 \\		
			$y_{it} = k$ & $c_{itk}$  & 0 & 1       & 2            &\dots           & 31    &32             \\ \hline
			0            & 0 & 1 & 0       & 0     & \dots       & 0            & 0                   \\
			0 & 1 & 0 & 0       & 0     & \dots                    &0     & 0        \\
			& & & & & & & \\
			1            & 0 & 0 & $1-p_1$ & 0           & \dots       & 0            & 0                   \\
			1 & 1 & 0 & $p_1$   & $p_1$                   & \dots     & $p_1$        & $p_1$               \\
			& & & & & & & \\
			2            & 0 & 0 & 0       & $(1-p_1)(1-p_2)$    & \dots      & 0         & 0                   \\
			2 & 1 & 0 & 0       & $(1-p_1)p_2$      & \dots        & $(1-p_1)p_2$     & $(1-p_1)p_2$        \\
			& & & & & & & \\
			\vdots            & \vdots & \vdots & \vdots       & \vdots     & $\ddots$& \vdots       & \vdots            \\
			& & & & & & & \\
			31            & 0 & 0 & 0       & 0                     & \dots        & $\prod_{i=1}^{31} (1-p_i)$        & 0          \\
			31 & 1 & 0 & 0       & 0    &\dots                     & $\prod_{i=1}^{30} (1-p_i)p_{31}$       & $\prod_{i=1}^{30} (1-p_i)p_{31}$             \\
			& & & & & & & \\
			32            & 0 & 0 & 0       & 0              & \dots        & 0      & 0            \\
			32 & 1 & 0 & 0       & 0           & \dots       & 0      & $\prod_{i=1}^{31} (1-p_i)p_{32}$                    \\\hline
		\end{tabular}
	\end{center}
\end{table}

Although the probability $\Pr(Z_{it} = \ell)$ seems to follow a known distribution, it does not follow a smooth distribution, see Figure \ref{fig:numCards}. The left panel shows the number of cards a child intends to turn over, that is, proportionally the empirical equivalence of $\Pr(Y_{it} = k \mid C_{itk} = 0) = \Pr(Z_{it} = k)$. It is easy to see that some outcomes seem extra attractive. We choose to distinguish four of these cases: (a) $k = 0$, (b) $k \in A$ with $A = \{4, 8, 10, 12, 16, 20, 24\}$, (c)  $k = 31$, and (d) otherwise. To control for these four cases, we implement a multiple inflation model, where the observations belonging to each of these cases get extra probability mass through parameter $\phi_m$ with $m = 1, \dots, 4$. The probability $\Pr(Z_{it} = \ell)$ is defined by
\begin{equation*}
\Pr(Z_{it} = \ell) = \left\{
\begin{array}{ll}
\phi_1 f(0) + \phi_2  & \text{if } \ell = 0,\\
\phi_1 f(\ell) + \frac{1}{|A|} \phi_3 & \text{if } \ell \in A,\\
\phi_1f(31) + \phi_4  & \text{if } \ell = 31,\\
\phi_1 F(32) & \text{if } \ell = 32,\\
\phi_1 f(\ell) & \text{for all other values } \ell,
\end{array} 
\right.
\end{equation*}
where $f(\ell)$ and $F(\ell)$ are the probability mass function and cumulative distribution function, respectively, of a known distribution. Note that the weights $\phi_m$ have to be between zero and one, $0 \leq \phi_m \leq 1$, and sum to one, $\sum_{m=1}^{4} \phi_m = 1$. 

Since the number of cards someone intends to turn over ($z_{it}$) is nonnegative and discrete, the distribution has to have these properties as well. We choose the negative binomial distribution, because it allows the variance to differ from the mean, in contrast to the Poisson distribution. The negative binomial distribution can be written as a Poisson-Gamma mixture. Specify the mean of the Poisson distribution as a combination of a deterministic function of the predictors, $\mu_{it} = g(\eta_{it})$, and a random component, $\nu \sim_{\text{i.i.d.}}g(\nu \mid \kappa)$. Let $g(\nu \mid \kappa)$ be the density of the Gamma distribution, then the resulting Poisson-Gamma mixture density can be rewritten as the negative binomial density. The probability mass function of this distribution is specified as follows
\begin{equation*}
f(z_{it} \mid \mu_{it}, \delta) = \frac{\Gamma (\delta + z_{it})}{\Gamma(\delta) z_{it}!} \left(\frac{\delta}{\delta + \mu_{it}}\right)^{\delta} \left(\frac{\mu_{it}}{\mu_{it} + \delta}\right)^{z_{it}},
\end{equation*}
where $\delta = 1/\kappa$.

\begin{figure}
	\centering
	\includegraphics[width=0.5\linewidth]{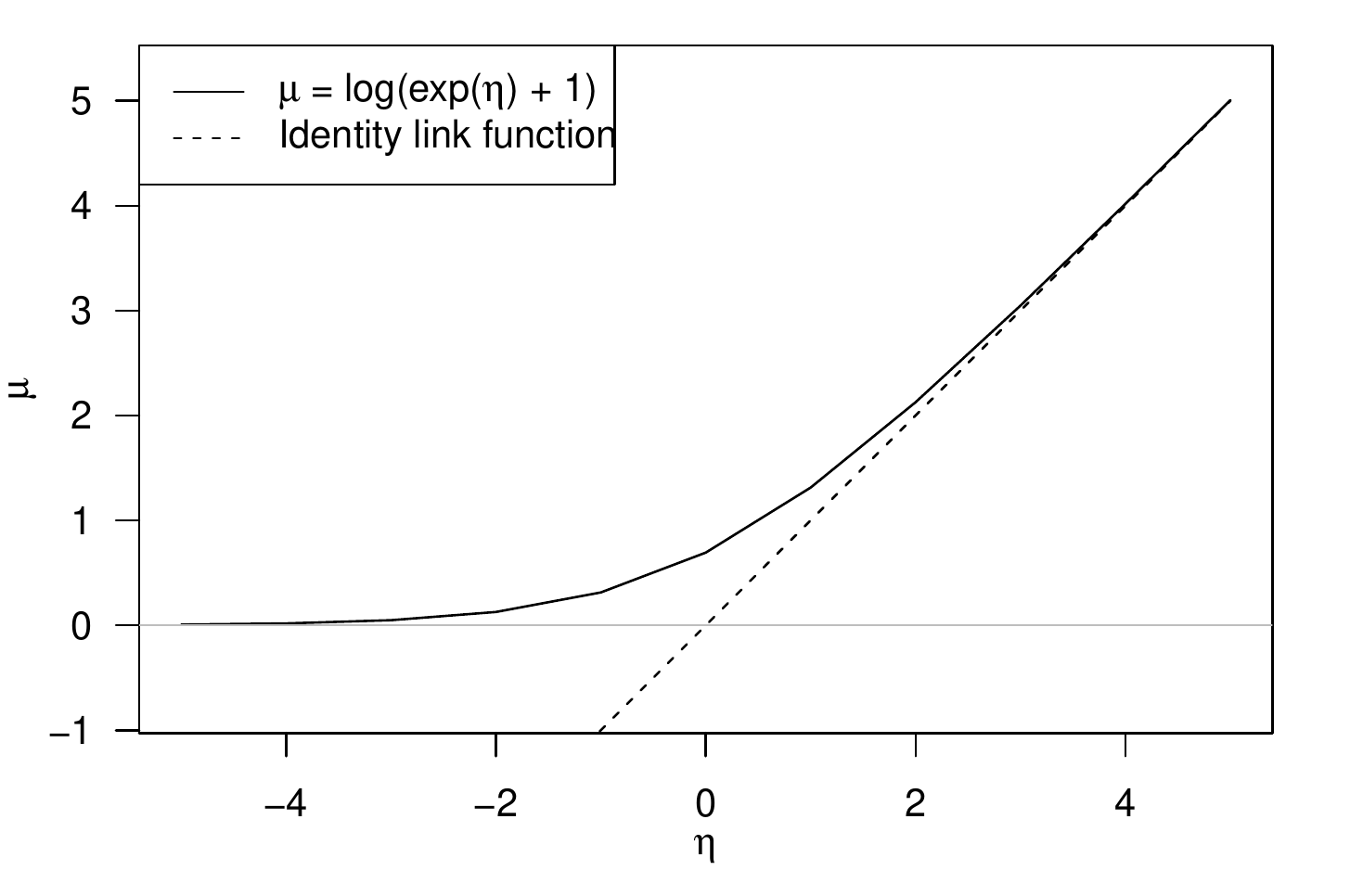}
	\caption{Both the proposed inverse link function, $\mu_{it} = h^{-1}(\eta_{it}) = \log(\exp(\eta_{it}) +1)$, and the identity link function, $\mu_{it} = h^{-1}(\eta_{it}) = \eta_{it}$, on the domain [-5:5].}
	\label{fig:link}
\end{figure}

In a generalized linear model (GLM), the mean $\mu_{it}$ of the distribution is specified through a inverse link function
\begin{equation*}
\mu_{it} = h^{-1}(\eta_{it}).
\end{equation*}
The mean $\mu_{it}$ of the negative binomial distribution must be larger than zero. Therefore, the inverse link function $ h^{-1}(\eta_{it})$ should map $\eta_{it} \in \mathbb{R}$ to  $\mathbb{R}^+$. In GLM, $\eta_{it}$ is chosen as a linear combination of covariates $\bm{x}'_{it}$, that is, 
\begin{eqnarray*}
	\eta_{it} = \alpha + \bm{x}'_{it} \bm{\beta}.
\end{eqnarray*}
For ease of interpretation of the coefficients $\bm{\beta}$,  we specify, the inverse link function by
\begin{equation*}
\label{eq:mu}
h^{-1}(\eta_{it}) = \log(\exp(\eta_{it}) +1)
\end{equation*}
so that whenever $\eta_{it} > 1$, the inverse link function becomes close to linear, yet $h^{-1}(\eta_{it}) > 0$ for any $\eta_{it}$, see Figure \ref{fig:link} \cite<see, e.g.,>{ranganath2016hierarchical}.

The predictor variables are all gathered in the vector $\bm{x}_{it}$. Some predictor variables are categorical and we choose to represent each of them by their own dummy variable. Without loss of generality, the weights corresponding to the dummy variables for each categorical variable must have sum zero, that is, 
\begin{eqnarray*}
	\bm{C\beta} = \bm{0} \mbox{~with~} 
	\bm{C} = \left[\begin{array}{ccc}
		(\bm{1'1})^{-1/2}\bm{1}' & \bm{0}' & \bm{0}' \\
		\bm{0}' & (\bm{1'1})^{-1/2}\bm{1}' & \bm{0}' \\
	\end{array}\right],
\end{eqnarray*}
where, for illustration, it is assumed that there are two categorical variables followed by numerical predictors, so that  $\bm{C}$ consists of one row per categorical variable with ones at the positions of the weights and zero elsewhere. Note that the factors $(\bm{1'1})^{-1/2}$ are chosen for notational convenience. Again, without loss of generality, it is also assumed that the numerical predictor variables are $z$-scores (with mean zero and standard deviation one) so that the intercept $\alpha$ can be interpreted as an overall measure of risk taking for someone who has a neutral score on all predictors. 

The complete likelihood over all $N$ individuals becomes
\begin{eqnarray*}
	L(\bm{\theta}) &=& \prod_{i = 1}^{N} L_i = \prod_{i = 1}^{N} \prod_{t = 1}^{T} \Omega_{y_{it}z_{it}, c_{ity_{it}}} \Theta_{z_{it}, c_{ity_{it}}} \\
	&=&  \left({\prod_{i = 1}^{N} \prod_{t = 1}^{T} \Omega_{y_{it}z_{it}, c_{ity_{it}}} }\right)  
	\left({\prod_{i = 1}^{N} \prod_{t = 1}^{T} \Theta_{z_{it}, c_{ity_{it}}} }\right)\\
	&\propto& \prod_{i = 1}^{N} \prod_{t = 1}^{T} \Theta_{z_{it}, c_{ity_{it}}}, 
\end{eqnarray*} 
where $\bm{\theta}$ is the vector of all unknown parameters. The factor $\prod_{i = 1}^{N} \prod_{t = 1}^{T} \Omega_{y_{it}z_{it}, c_{ity_{it}}} $ is irrelevant for maximizing the likelihood as it is constant, so that optimizing  $\prod_{i = 1}^{N} \prod_{t = 1}^{T} \Theta_{z_{it}, c_{ity_{it}}}$ over $\bm{\theta}$ is sufficient.

In a final step of the CMM, we wish to be able to model unobserved heterogeneity across individuals by adding finite mixtures with different intercepts per segment to the model, that is, 
\begin{equation*}
\eta_{its} = \alpha_s + \bm{x}'_{it}\bm{\beta},
\end{equation*}
with $\alpha_s$ the segment specific intercept. The relative size of the segment is estimated by  $\pi_s$. Then, the likelihood function becomes
\begin{eqnarray}
L(\bm{\theta}) & = &\prod_{i = 1}^{N} \sum_{s = 1}^{S} \pi_s \prod_{t = 1}^{T} \Omega_{y_{it}z_{it}, c_{ity_{it}}}  \prod_{t = 1}^{T} \Theta_{z_{it}, c_{ity_{it}},s}\nonumber \\
& = &\left({\prod_{i = 1}^{N} \prod_{t = 1}^{T} \Omega_{y_{it}z_{it}, c_{ity_{it}}}  }\right) 
\left({\prod_{i = 1}^{N} \sum_{s = 1}^{S} \pi_s \prod_{t = 1}^{T} \Theta_{z_{it}, c_{ity_{it}},s} }\right) \nonumber \\
& \propto &\prod_{i = 1}^{N} \sum_{s = 1}^{S} \pi_s \prod_{t = 1}^{T} \Theta_{z_{it}, c_{ity_{it}},s}, \label{eq:ll}
\end{eqnarray}
where $\bm{\theta}$ is understood to contain all unknown parameters. Note that $\Theta_{z_{it}, c_{ity_{it}},s}$ has obtained an additional subscript $s$ to indicate that this probability is dependent on the parameter $\alpha_s$. Thus, the CMM needs to maximize $L(\bm{\theta})$ over $\bm{\theta} '=  [\bm{\alpha}', \bm{\beta}', \delta, \bm{\phi}', \bm{\pi}']$ subject to $\bm{C\beta} = \bm{0}$, $\phi_m \geq 0$, $\bm{1'\phi} = 1$, $\pi_s \geq 0$, and $\bm{1'\pi} = 1$. More details about the estimation procedure can be found in  Appendix~\ref{sect:MaxLik}.

\section{Results}
Before the Censored Mixture Model (CMM) can be applied to the data discussed in Section~\ref{sect:data}, several parameters need to be set. First, the maximization of the log likelihood function is performed through the \texttt{optimx} function in \texttt{optimx} package in \texttt{R}. All default settings are used except for the relative convergence tolerance, \textit{reltol},  which is set more strictly such that the maximization has converged as soon as $\log L(\bm{\theta}^{(t)})- \log L(\bm{\theta}^{(t-1)})$ is less than $10^{-10} (\mid \log L(\bm{\theta}^{(t)}) \mid + 10^{-10})$ where $t$ is the iteration counter. After convergence, one step of the Newton-Raphson method is performed using a numerically approximated Hessian with the aim of ensuring that the gradient is close to zero. To speed up the convergence, the start values of $\bm{\alpha}$ and $\bm{\phi}$ are based on educated guesses. For $\bm{\alpha}$, the start values are uniformly distributed over the possible outcomes, $\{0, 32\}$. The initial values of $\bm{\phi}$ are based on the observed proportion of excesses in Figure \ref{fig:numCards}, that is, the difference of the observed proportion of the inflated outcome minus the interpolated value of the previous and next outcomes without inflation. To further improve the convergence speed, we trained our model on a small subsample, $n = 100$, and implemented these parameter estimates as start values of $\bm{\theta}$ in the model using the original sample.

\subsection{Selection of the Number of Segments}
\begin{figure}
	\centering
	\includegraphics[width=0.6\linewidth]{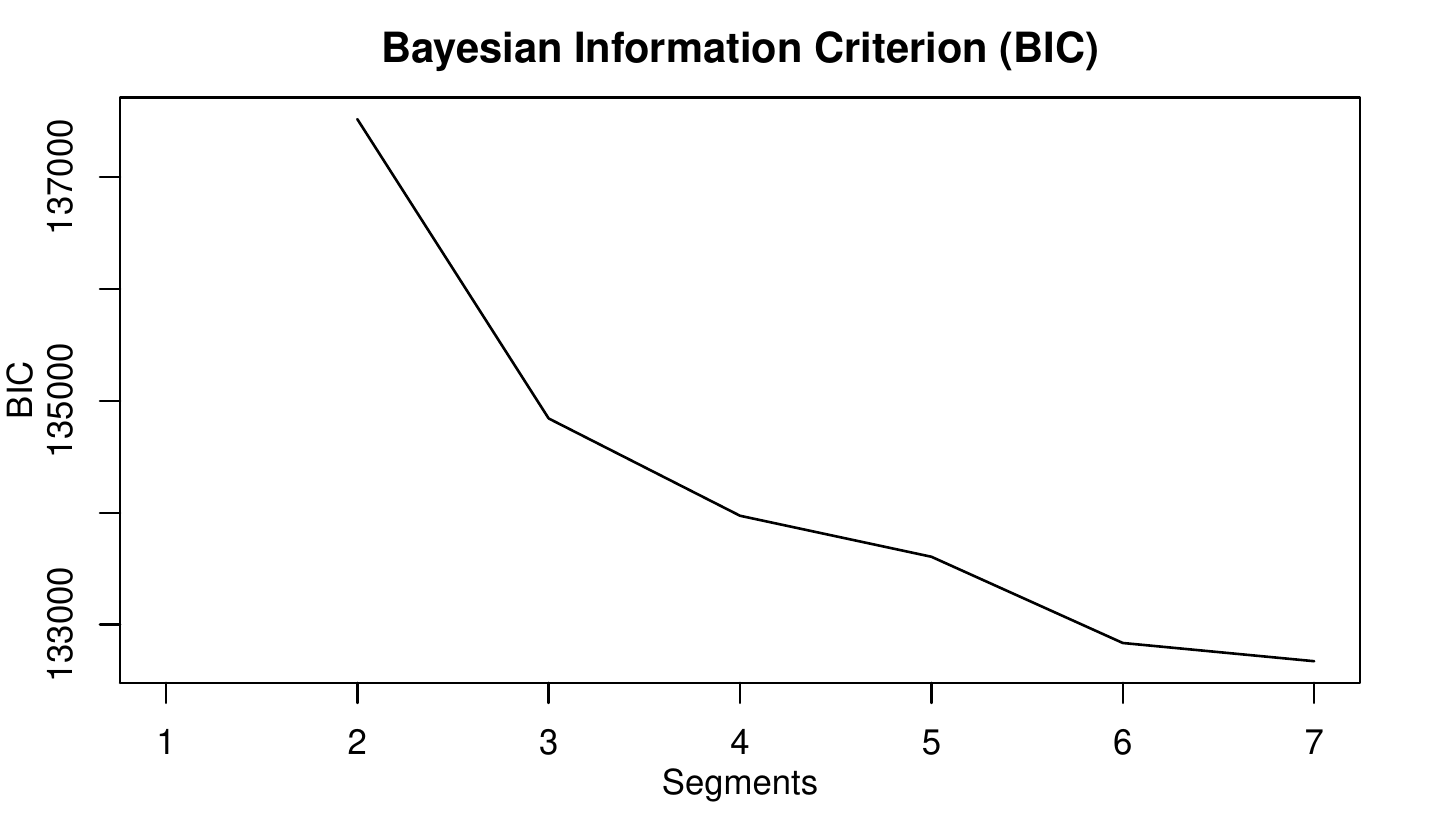}
	\caption{The Bayesian Information Criterion (BIC) of CMMs with $S = 2$ to 7 segments.}
	\label{fig:bic}
\end{figure}

As the number of segments is unknown a priori, the model is computed for several numbers of segments $S$. We use several criteria to decide on a useful number of segments: the Bayesian Information Criterium (BIC), a minimum segment size, and the distinctiveness of the segment specific intercept. The BIC for various choices of $S$ is shown in Figure~\ref{fig:bic}. Since the number of observations is so large in this study, adding a segment hardly affects the BIC. Therefore, searching for the number of segments that would lead to a minimum BIC would require an unrealistically high number of segments. Therefore, we additionally check the size of the segments $\pi_s$ and the segment specific intercepts $\alpha_s$ given in Table~\ref{tab:seg}. We opt for segments that have $\pi_s > 5\%$ of the observations, that is, 170 children. Furthermore, we impose the segment specific intercepts $\alpha_s$ to be sufficiently different to avoid segments where the level of risk seeking as symbolized by their respective $\alpha_s$ is hardly different. Based on these three criteria, we choose to continue interpreting the model with $S = 4$ segments.

From Table \ref{tab:seg} it is clear that Segment 1 is overall the smallest ($\pi_1 = 0.097$) and is characterized by children that are on average most risk averse as $\alpha_1$ is the smallest of all segments. In contrast, the last segment contains children that are most risk seeking as their intercept $\alpha_4 = 37.52$ is even larger than the total number of cards that could be turned over in the game.

\begin{table}
	\centering
	\caption{Segment probabilities $\pi_s$ and segment specific intercepts $\alpha_s$ with the standard errors between brackets for CMMs with $S = 2$ to 7 segments.}
	\label{tab:seg}
	\begin{threeparttable}
		\begin{tabular}{llrrrrrrr}
			\hline
			  & &   \multicolumn{7}{c}{Segment $s$}  \\ \cline{3-9}
			$S$ && 1~~    & 2~~     & 3~~     & 4~~     & 5~~     & 6~~     & 7~~ \\ \hline
			2    &  &       &       &       &       &       &       &  \\
			&$\pi_s$    & 0.394 &  0.606 &   &       &       &       &  \\
			& & (0.010) & (0.010) & & & & & \\
			&$\alpha_s$ & 9.90 & 26.35 &  &       &       &       &  \\
			& & (0.164) & (0.313) & & & & & \\
			3    &  &       &       &       &       &       &       &  \\
			&$\pi_s$    & 0.150 &  0.432 &  0.419 &       &       &       &  \\
			& & (0.007) & (0.010) & (0.011) & & & & \\
			&$\alpha_s$ & 6.77 & 13.93 & 30.93 &       &       &       &  \\
			& & (0.148) & (0.184) & (0.408) & & & & \\
			4 &    &       &       &       &       &       &       &  \\
			& $\pi_s$    & 0.097 &  0.275 &  0.357 &  0.271 &       &       &  \\
			& & (0.006) & (0.011) & (0.012) & (0.011) & & & \\
			& $\alpha_s$ & 5.85 & 11.04 & 18.68 & 37.52 &       &       &  \\
			& & (0.152) & (0.188) & (0.295) & (0.772) & & & \\
			5 &    &       &       &       &       &       &       &  \\
			&$\pi_s$    & 0.023 &  0.119 &  0.284 &  0.331 &  0.243 &       &  \\
			& & (0.003) & (0.007) & (0.011) & (0.012) & (0.012) & & \\
			& $\alpha_s$ & 3.11 & 6.89 & 11.74 & 19.44 &  38.40 &       &  \\
			& & (0.167) & (0.148) & (0.187) & (0.322) & (0.904) & & \\
			6 &    &       &       &       &       &       &       &  \\
			&$\pi_s$    & 0.002 &  0.103 &  0.206 &  0.256 &  0.249 &  0.164 &  \\
			& & (0.003) & (0.008) & (0.021) & (0.015) & (0.017) & (0.015) & \\
			&$\alpha_s$ & 2.67 &  6.55 & 10.64 & 15.58 & 23.97 & 46.82 &  \\
			& & (0.258) & (0.168) & (0.315) & (0.587) & (0.852) & (2.685) & \\
			7 &    &       &       &       &       &       &       &  \\
			&$\pi_s$    & 0.007 &  0.052 &  0.130 &  0.294 &  0.303 &  0.000 &  0.214 \\
			& & (0.002) & (0.005) & (0.009) & (0.012) & (0.012) & (0.000) & (0.013) \\
			& $\alpha_s$ & -0.66 & 5.03 & 8.36 & 12.99 & 21.07 & 22.70 & 40.83 \\
			& & (0.225) & (0.163) & (0.188) & (0.225) & (0.438) & (148.3) & (1.328) \\ \hline
		\end{tabular}%
	\end{threeparttable}
\end{table}%

\subsection{Segments Specific Results}

\begin{figure}
	\centering
	\includegraphics[width=0.6\linewidth]{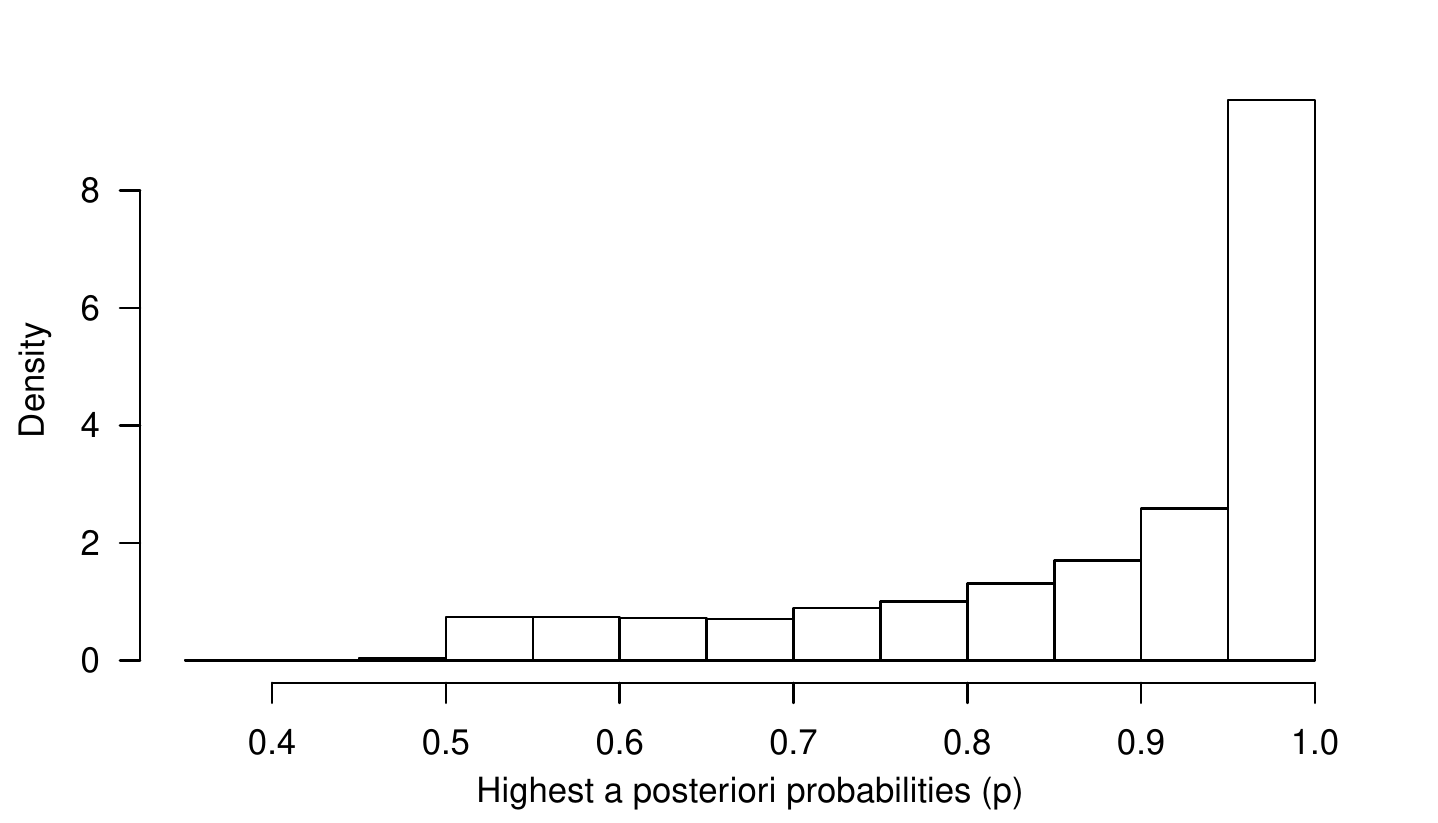}
	\caption{A histogram of the highest a posteriori segment probability of each individual.}
	\label{fig:segProb}
\end{figure}

For each individual, we can compute the a posteriori probability of belonging to a segment. Ideally these probabilities are close to one for one of the segments and close to zero for the others thereby clearly assigning an individual to a segment. To see how distinctive the segments are, we consider the highest a posteriori probability for each individual and plot that in a histogram. Figure~\ref{fig:segProb} shows this distribution and it is clear that indeed most children are assigned to a segment with a large probability. Therefore, virtually each child can be assigned with high probability to one of the segments. 

It is interesting to investigate how the segments differ on characteristics that have not been part of the model. In particular, how do the segments differ with regard to the occurrence of behavioral problems as measured by the CBCL? The resulting z-scores (with mean zero and standard deviation one) per segment weighted by the a posteriori probabilities per segment are presented in Table~\ref{tab:segChar}. Appendix~\ref{sect:sigtest} discusses how to test for differences of weighted means. The stars in Table~\ref{tab:segChar} denote whether one of the segment averages is significantly different from the overall average for this particular symptom.

The level of behavioral problems in all subscales except that of social withdrawal differ between the groups of children as defined by our segments. The risk averse children in Segment 1 and the risk seekers in Segment 4 on average have more behavioral problems than the children in Segment 2 and 3. Table~\ref{tab:segChar} also suggests that children in Segment 3 who intend to turn over on average 18	 cards score on average the lowest on all CBCL subscales. Furthermore, we can see from this table that a risk averse strategy is most profitable, as the average score is highest in Segment 1 (most risk averse segment) and lowest in Segment 4 (most risk seeking segment).

\begin{table}[htbp]
	\centering
	\caption{Weighted z-scores per segment of CBCL subscales scores and other CCT characteristics.}
	\begin{threeparttable}
	\begin{tabular}{>{\quad}lrrrrr}
		\hline
		& \multicolumn{5}{c}{Segment $s$} \\
		\cline{2-6}
		& \multicolumn{1}{c}{1} & \multicolumn{1}{c}{2} & \multicolumn{1}{c}{3} & \multicolumn{1}{c}{4} & \multicolumn{1}{c}{Total} \\ \hline
		\rowgroup $\pi_s$    & 0.10  & 0.28  & 0.36  & 0.27  &  \\
		&       &       &       &       &  \\
		\rowgroup CBCL subscales &       &       &       &       &  \\
		Internalizing \tnote{*} & 0.06  & 0.00  & -0.05 & 0.05  & 0.00 \\
		Externalizing \tnote{**} & 0.08  & -0.03 & -0.04 & 0.06  & 0.00 \\
		\rowgroup CBCL symptom subscales &       &       &       &       &  \\
		Anxiety \tnote{*} & 0.07  & 0.01  & -0.06 & 0.04  & 0.00 \\
		Social withdrawal \tnote{*} & 0.02  & 0.02  & -0.06 & 0.05  & 0.00 \\
		Somatic complaints & 0.03  & -0.02  & -0.01 & 0.02  & 0.00 \\
		Social problems \tnote{***} & 0.09  & -0.04 & -0.05 & 0.08  & 0.00 \\
		Thought problems \tnote{**} & 0.10  & -0.02  & -0.05 & 0.05  & 0.00 \\
		Attention problems \tnote{***} & -0.03  & -0.04 & -0.06 & 0.13  & 0.00 \\
		Delinquent behavior \tnote{**} & 0.00  & -0.03 & -0.03 & 0.08  & 0.00 \\
		Aggressive behavior \tnote{*} & 0.10  & -0.02  & -0.04 & 0.04  & 0.00 \\
		&       &       &       &       &  \\
		\rowgroup Average score \tnote{***} & -87.0 & -123.7 & -170.7 & -230.1 & -165.7 \\
		\rowgroup $\#$ cards turned over \tnote{***} & 5.0   & 7.7   & 10.0   & 11.7  & 9.3 \\
		\rowgroup $\#$ censored trials \tnote{***} & 5.8   & 8.5   & 11.4  & 14.3  & 10.8 \\ \hline
	\end{tabular}%
		A Wald test is performed to check for a significant difference between the segments. One star denotes $0.05 \leq p <0.10$,  two $0.01 \leq p <0.05$, and three $p<0.01$. The 223 children without a CBCL score measured at either six or nine years old were excluded.
	\end{threeparttable}
	\label{tab:segChar}%
\end{table}%

One of the contributions of the CMM model is that the intended number of cards to be turned over is estimated by the segment specific intercept $\alpha_s$. Due to the censoring, children who intend to turn over a high number of cards will often not be able to do so. Therefore, the observed number of cards turned over under estimates the intended number of cards to be turned over. We can easily compare them using the forelast row in Table~\ref{tab:segChar} with the $\alpha_s$ from Table~\ref{tab:seg}. For example, the average number of observed cards turned over by children in Segment 1 is 5.0 whereas average number of card intended is 5.9. For Segments 2, 3, and 4, these values are 7.7, 10.0, and 11.7 observed and 11.0, 18.7, and 37.5 intended. Indeed, a large under estimation of risk seeking is obtained when only considering the observed number of cards turned over.

\subsection{Regression Coefficients}

The regression coefficients $\bm{\beta}$ are presented in Table \ref{tab:beta}. The two numerical variables (age and IQ) are standardized to $z$-scores prior to the analysis. Whether or not a loss card was drawn in the last two games is recorded by the following variables: previous loss yes, previous loss no, second previous loss yes, and second previous loss no. As our link function in Figure~\ref{fig:link} is close to the identity function for values larger than 1, the coefficients can be interpreted on the scale of the number of cards turned over. For categorical variables, we chose mean  weights of the categories belonging to a single variable to be zero so that the intercept can be interpreted as the average score in the segment for a neutral child. As a consequence, the difference in weights between two categories is the corresponding effect, for example, girls on average turn over $0.286 + \mid-0.286\mid =  0.572$ cards more than boys.

Furthermore, age and IQ have a negative association with the number of cards turned over. Also, a higher household income is related to higher levels of risk taking. Children with a mother with a Dutch or Asian ethnicity turn over fewer cards than the base average.

Due to the different game settings, we are able to investigate the effect of the loss probability and the sensitivity to reward and punishment. According to the model, the number of loss cards has the strongest effect on the number of cards turned over. In a game with three loss cards on average 1.7 cards less are turned over, than in a game with one loss card. The game setting loss amount also shows the expected direction of effect. In a trial with a high loss amount the expected number of cards turned over is lower than in a trial with a low loss amount. Unexpectedly, the predicted number of cards turned over is lower in a trial with a high gain amount than it is in a trial with a low gain amount.

Moreover, the results in the previous round have a strong association with observed behavior in the current round. If a loss card was encountered in the previous round, on average 1.6 cards less are turned over. The experience of a loss card two trials earlier also relates to the intention to turn over one card less in the current trial.

We included interaction terms between the game settings and sex. According to the estimates, the combination of boy and a loss amount of 250 accounts for an additional 0.063 $(= -0.286 + 0.195 + 0.154)$ cards to be turned over. In a trial with loss amount 750, a boy is expected to turn over 0.635 $(= \mid -0.286 -0.195 -0.154\mid)$ cards less than the base average (i.e., the segment specific intercepts). Hence, the effect the loss amount has on the number of cards turned over by boys is 0.698 ($=0.063 + \mid -0.635 \mid$). This effect is smaller for girls, namely 0.572 ($=(0.286 +0.195 -0.154) + (0.286 - 0.195 + 0.154)$). Therefore, boys seem to be more sensitive to punishment in the CCT than girls are. Moreover, boys are also more influenced by the number of loss cards (2.038 vs. 1.36), whereas girls seem to be more sensitive to reward than boys are (0.346 vs. 1.026).

\begin{landscape}
\begin{table}
	\centering
	\caption{Regression coefficients with their standard errors. Within a categorical variable the sum of coefficients sum to zero and the continues variables age and IQ are standardized.}
	\begin{threeparttable}
		\begin{tabular}{>{\quad}lrrrlrr}
		\hline
		\rowgroup Background variables & \multicolumn{2}{c}{\begin{tabular}[c]{@{}c@{}}$\beta$ - coefficients\\ (st error)\end{tabular}} &      & \rowgroup Game settings      & \multicolumn{2}{c}{\begin{tabular}[c]{@{}c@{}}$\beta$ - coefficients\\ (st error)\end{tabular}}\\ \hline
		\rowgroup Age   & -0.012 & (0.071) &       & \rowgroup Gain amount (10) & 0.343 & (0.039) \\
		\rowgroup Boy   & -0.286 & (0.079) &       & \rowgroup Gain amount (30) & -0.343 & (0.039) \\
		\rowgroup Girl  & 0.286 & (0.079) &       & \rowgroup Loss amount (250) & 0.195 & (0.038) \\
		\rowgroup IQ    & -0.539 & (0.095) &       & \rowgroup Loss amount (750) & -0.195 & (0.038) \\
		\rowgroup Ethnicity mother &       &       &       & \rowgroup Loss cards (1) & 0.850 & (0.039) \\
		Dutch & -1.170 & (0.157) &       & \rowgroup Loss cards (3) & -0.850 & (0.039) \\
		Asian & -0.875 & (0.258) &       & \rowgroup Previous loss yes & -0.823 & (0.040) \\
		African & 0.570 & (0.393) &       & \rowgroup Previous loss no & 0.823 & (0.040) \\
		Moroccan & 0.477 & (0.338) &       &  \rowgroup Second previous loss yes & -0.502 & (0.040) \\
		Dutch Antilles & -0.139 & (0.379) &       & \rowgroup Second previous loss no & 0.502 & (0.040) \\
		Surinamese & 0.288 & (0.378) &       & \rowgroup Interaction terms &       &  \\
	    Turkish & 0.527 & (0.342) &       & Gain amount (10) : Boy & -0.170 & (0.039) \\
		Other Western & 0.322 & (0.265) &       & Gain amount (30) : Boy & 0.170 & (0.039) \\
		\rowgroup Education mother &       &       &       & Gain amount (10) : Girl & 0.170 & (0.039) \\
		No or primary education & 0.571 & (0.219) &       & Gain amount (30) : Girl & -0.170 & (0.039) \\
		Secondary education & -0.231 & (0.143) &       & Loss amount (250) : Boy & 0.154 & (0.038) \\
		Higher education & -0.340 & (0.137) &       & Loss amount (750) : Boy & -0.154 & (0.038) \\
		\rowgroup Householdincome per month in euro's &       &       & & Loss amount (250) : Girl & -0.154 & (0.038) \\
		$< 2000$ & -0.134 & (0.163) &       &  Loss amount (750) : Girl & 0.154 & (0.038) \\
		$2000 - 4000$ & -0.231 & (0.114) &      & Loss cards (1) : Boy & 0.169 & (0.039) \\
		$> 4000$ & 0.365 & (0.123) &       & Loss cards (3) : Boy & -0.169 & (0.039) \\
		&&&& Loss cards (1) : Girl & -0.169 & (0.039) \\
		&&&& Loss cards (3) : Girl & 0.169 & (0.039) \\ \hline
	\end{tabular}%
	\end{threeparttable}
	\label{tab:beta}%
\end{table}%
\end{landscape}

\subsection{Model Performance}
Our model gives of each child on each trial a probability distribution for the number of cards turned over. To obtain a sense how well the model fits the observed uncensored number of cards turned over, we compute a point estimate as the expected value of that distribution. Then, the model performance can be judged in terms of the difference between observed and expected number of cards turned over. Appendix \ref{sec:pred} provides more details on how these expectations are computed. Predictions can be generated with our CMM. The in-sample root-mean-square-error (RMSE) is equal to 12.0 and the mean absolute deviation (MAD) of the residuals is equal to 6.0. On a scale of 0-32 cards that can be turned over, these average deviations seem reasonable. We can present the same measures for the test set. The out-of-sample RMSE is equal to 12.1 and the MAD is equal to 6.0, showing little difference between in-sample and out-of-sample accuracy.

Another way to evaluate the model performance is by comparing the distributions of the empirical and predicted number of cards turned over for the training and the test data, similar to Figure~\ref{fig:numCards}. We break down the comparison into a censored and uncensored cases. For a fair comparison between the observed and predicted number of cards turned over in the censored case, one has to multiply the distribution of the predicted outcome with the conditional probability of being censored. In case of one loss card this probability is independent of the number of cards turned over, that is,
\begin{equation*}
	p_k = \frac{1}{32} \ \ \ \ \text{ if $\#$loss cards } = 1.
\end{equation*}
Note that these probabilities are equal to the one described in Table \ref{tab:prob}. In case of three loss cards the probability of being censored is equal to
\begin{equation*}
\begin{aligned}
&p_1 = \frac{3}{32} && \text{ if $\#$loss cards } = 3\\
&p_2 = \frac{29}{32} \cdot \frac{3}{31} && \text{ if $\#$loss cards } = 3\\
&p_k = \frac{32-k}{32} \cdot \frac{32-k-1}{31} \cdot \frac{3}{30} && \text{ if $\#$loss cards } = 3 \cup k = 3, ..., 29.
\end{aligned}
\end{equation*}

\begin{figure}[]
	\centering
	\includegraphics[width=0.8\linewidth]{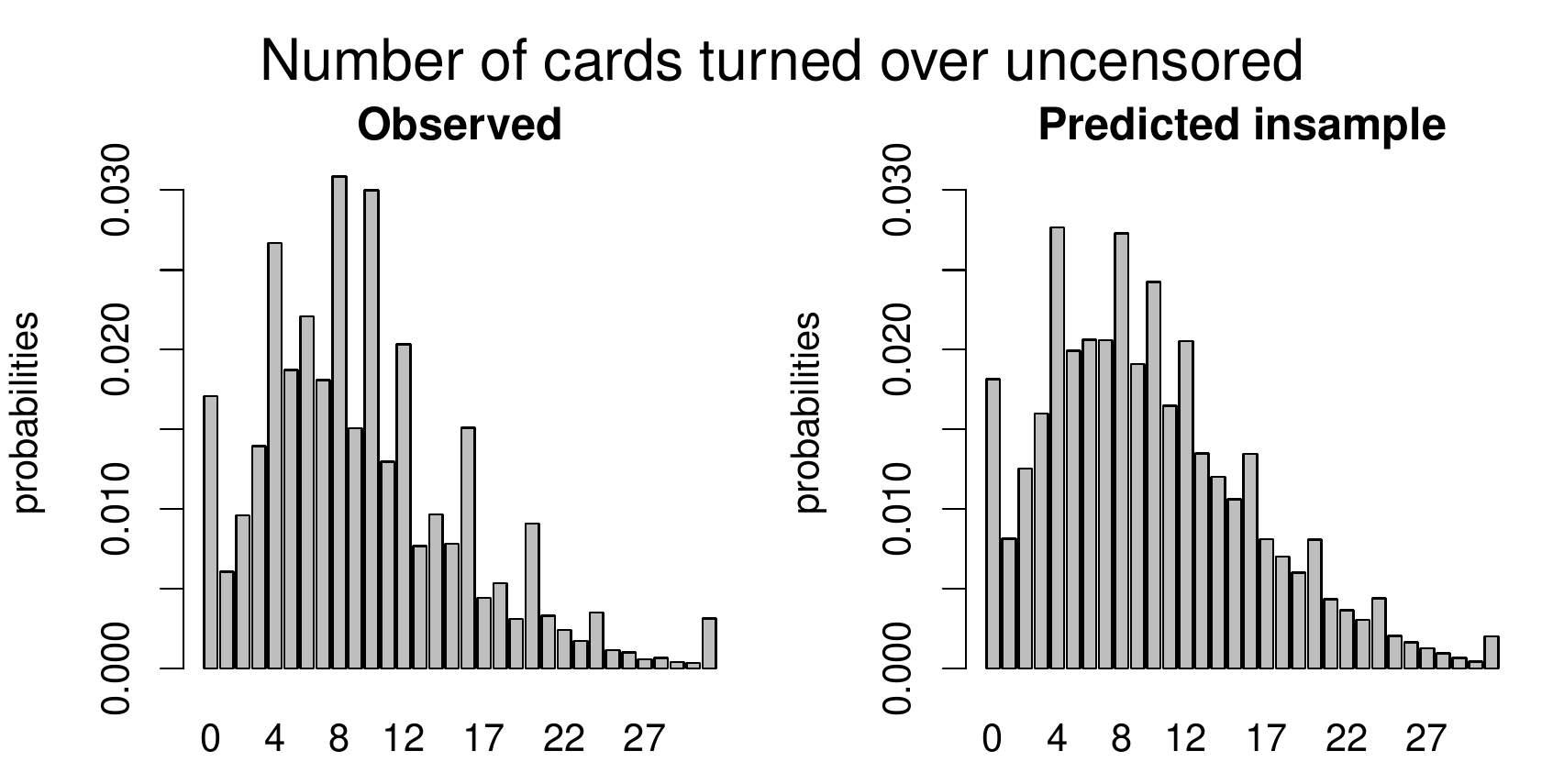}
	\caption{Distribution of the empirical (left panel) and predicted number of cards turned over by the CMM (right panel) for the uncensored observations in the training data.}
	\label{fig:uncen}
\end{figure}

\begin{figure}[]
	\centering
	\includegraphics[width=0.8\linewidth]{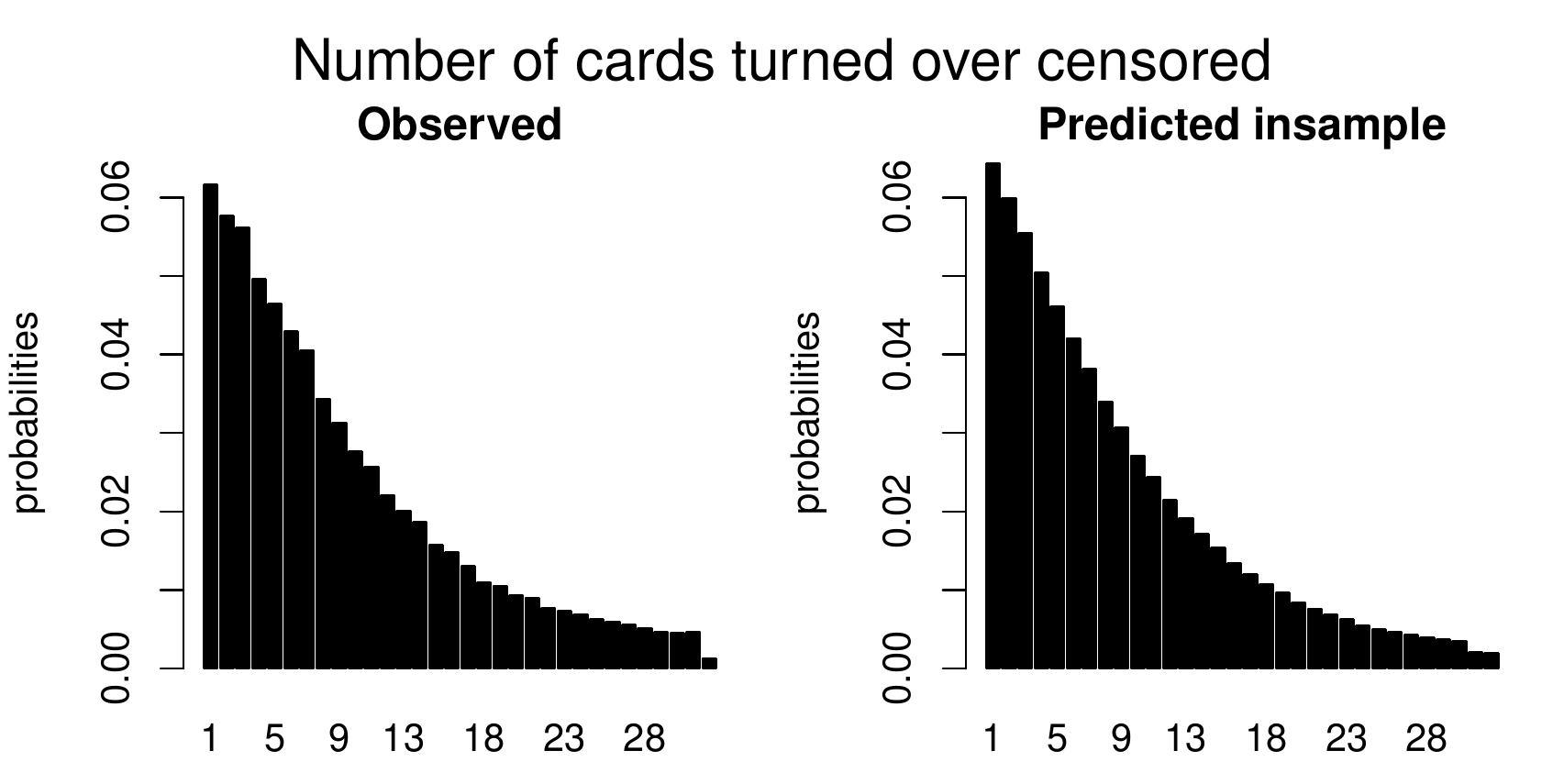}
	\caption{Distribution of the empirical (left panel) and predicted by the CMM (right panel) number of cards turned over corrected for the probability of being censored per card in the training data.}
	\label{fig:cen}
\end{figure}

\begin{figure}[]
	\centering
	\includegraphics[width=0.8\linewidth]{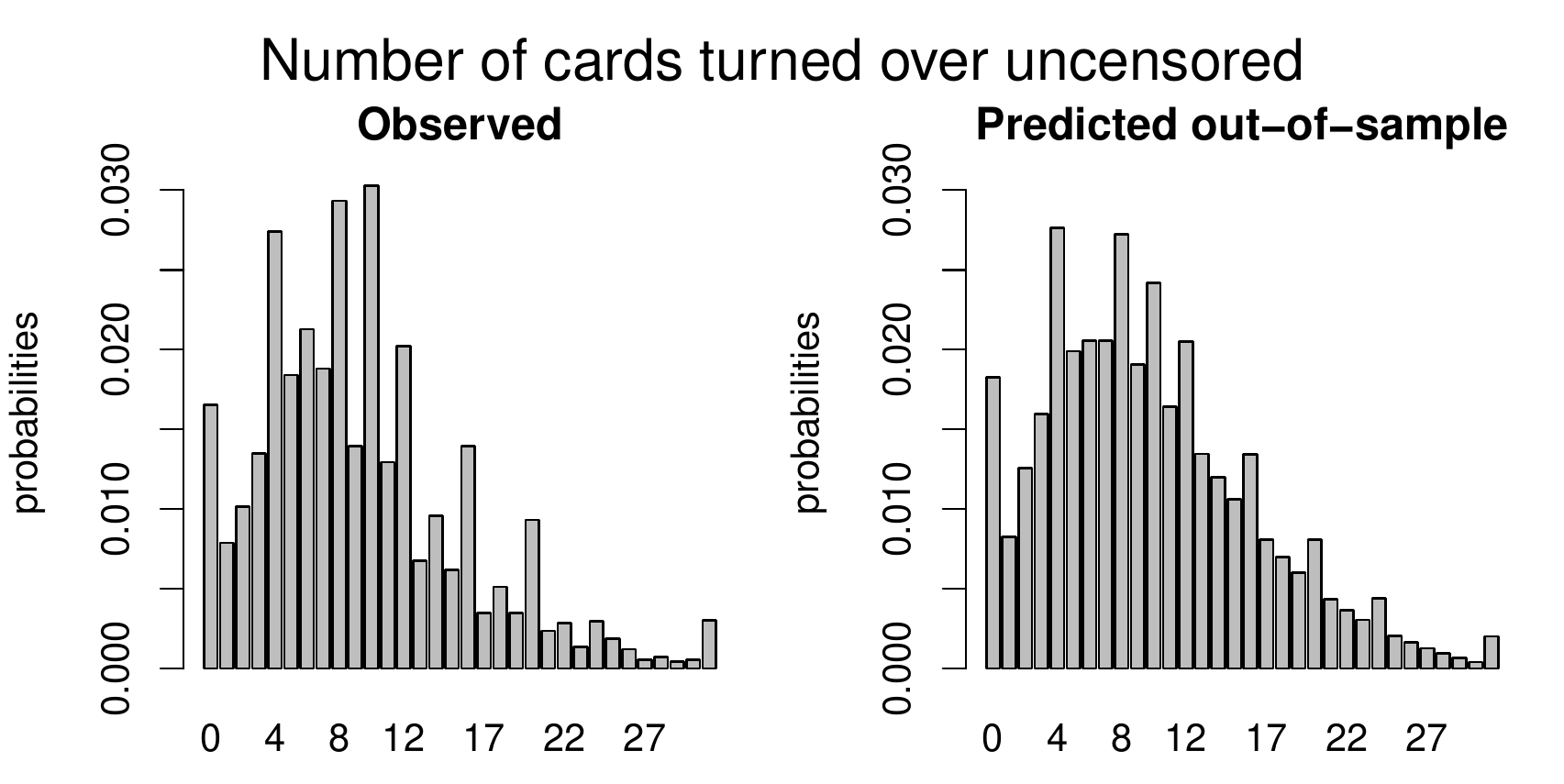}
	\caption{Distribution of the empirical (left panel) and predicted number of cards turned over by the CMM (right panel) for the uncensored observations in the test data.}
	\label{fig:uncenOofS}
\end{figure}

\begin{figure}[]
	\centering
	\includegraphics[width=0.8\linewidth]{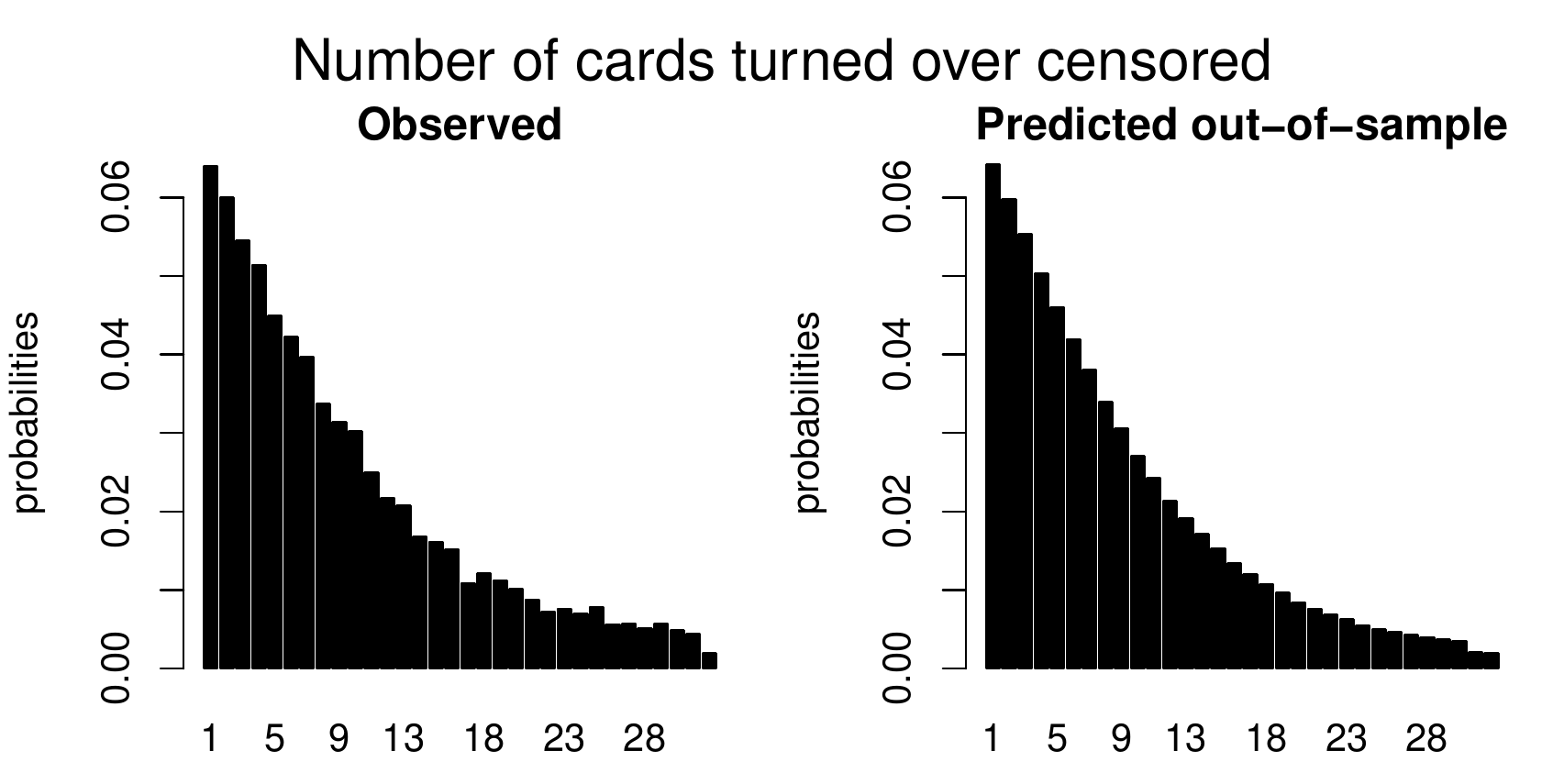}
	\caption{Distribution of the empirical (left panel) and predicted by the CMM (right panel) number of cards turned over corrected for the probability of being censored per card in the test data.}
	\label{fig:cenOofS}
\end{figure}

The empirical probability of the number of cards intended to turn over in the training set is shown in the left panel of Figure \ref{fig:uncen}. The comparison with the right panel with the CMM predicted probabilities shows that these predictions are quite accurate. The left panel of Figure~\ref{fig:cen} shows the empirical probability per card  of being censored in the training data and the right panel shows these values as predicted. Again, the distribution of the predicted values are similar to those observed. To guard against overfitting, we provide the same plots for the test set of 1049 children in Figures \ref{fig:uncenOofS} and \ref{fig:cenOofS}. The same interpretation holds as for the training data: there are some minor deviations from the observed distribution, but overall the test set prediction of these distributions is quite accurate.

\section{Discussion}
The Censored Mixture Model (CMM), developed to solve the potential problems emerging with modeling risk taking, is applied to an exceptionally large data set with 3404 children that each completed 16 rounds of the Columbia Card Task (CCT). To accommodate the potential censoring that often occurs in sequential risk tasks, the cumulative distribution function is added to the likelihood function. In the Generation R data set, the prevalence of censoring is $68\%$. Ignoring the censoring would seriously underestimate the intended level of risk taking as more than two third of the data would not be used. The inflated values in the outcome distribution are handled by assigning extra probability mass to these outcomes in the likelihood function. Figures \ref{fig:uncen}, \ref{fig:cen}, \ref{fig:uncenOofS}, and \ref{fig:cenOofS} clearly show peaks at certain outcome values in both the observed and predicted graphs. Without the extra probability mass for the inflated values, the distributions in the predicted graphs would have been smoother and, hence, less similar to the observed graphs. Finally, four mixtures, with a segment specific intercept are added to the model to account for unobserved heterogeneity across individuals. The distribution of posterior probabilities in Figure \ref{fig:segProb} clearly points out that the four segments are distinctive as large probabilities (say above $80\%$) are most prevalent. In case the individuals in the sample all have the same tendency for risk taking, this graph would be centered around 0.25, indicating that the individuals are assigned to all segments with equal probability.

The selection of the number of segments was based on a three way procedure. The BIC values of the different models are compared, the segment specific intercepts had to be distinctive among the segments, and, finally, the smallest segment had to contain at least five percent of the sample. Although we are very confident that a model with four segments is optimal in this case, a different strategy could have let to a different number of segments and, hence, slightly different results.

The CMM with four segments showed some interesting results. Both the most risk averse and risk seeking segments, respectively Segment 1 and 4, have the highest level of behavioral problems measured by the CBCL. Furthermore, children with a high IQ turn over fewer cards than the average and children from a family with a low household income turn over more cards than the average. Children with a Dutch, Asian, or Turkish background are more risk averse, whereas children with an African or Moroccan background are more risk seeking.


Moreover, from Table \ref{tab:beta} it is clear to see that the number of loss cards has the highest effect on the number of cards turned over, compared to the gain amount and loss amount. This result is in accordance with many other studies using the CCT \cite{kluwe2015assessing,holper2014hemodynamic,penolazzi2012impulsivity}. Looking at the risk neutral strategy based on the expected values (Table \ref{tab:ratCards}), it is clear that turning over zero cards is often most profitable. It would be interesting to see, whether different game settings show the same results. Note that we used the same game settings as \cite{figner2011takes}.

\begin{table}[]
	\centering
	\caption{Optimal number of cards to turn over when maximizing the expected value.}
	\label{tab:ratCards}
	\begin{tabular}{@{}llrrlllllllrrl@{}}
		\multicolumn{4}{c}{1 loss card}                                                                               &  &  &  &  &  & \multicolumn{4}{c}{3 loss cards}                                                                              &  \\ \cmidrule(r){1-4} \cmidrule(lr){10-13}
		&    & \multicolumn{2}{c}{loss amount} &  &  &  &  &  &                                                                        &    & \multicolumn{2}{c}{loss amount} &  \\
		&    & 250            & 750            &  &  &  &  &  &                                                                        &    & 250            & 750            &  \\ \cmidrule(r){1-4} \cmidrule(lr){10-13}
		\multirow{2}{*}{\begin{tabular}[c]{@{}l@{}}gain\\ amount\end{tabular}} & 10 & 7             & 0             &  &  &  &  &  & \multirow{2}{*}{\begin{tabular}[c]{@{}l@{}}gain\\ amount\end{tabular}} & 10 & 0             & 0              &  \\
		& 30 & 23             & 6             &  &  &  &  &  &                                                                        & 30 & 4              & 0             &  \\ \cmidrule(r){1-4} \cmidrule(lr){10-13}
	\end{tabular}
\end{table}

It would also be interesting for further research to collect more detailed information. These data could give us more insights in the underlying decision process. For example, the time between actions (turning over cards) can be used to investigate a potential fatigue effect. In addition, more time between actions at the end of a game round could indicate that a participant doubts between continuing and stopping. Moreover, the pattern and sequence of the cards turned over could strengthen our assumption that participants of the CCT create geometric patterns for turning over cards.

Finally, note that the negative binomial distribution has an infinite upper bound. This property implies that there is probability mass after 32, meaning that the model can predict to turn over more cards than possible. However, all the point estimates (for computations, see Appendix \ref{sec:pred}) are within the range $\{0, 32\}$, therefore we argue that this is not a major issue. In all, we believe that the Censored Mixture Model proposed in this paper is an important tool in the analysis of risk taking.

\pagebreak
\bibliographystyle{apacite}
\bibliography{mybib}

\pagebreak
\begin{appendices}
\appendix
\section{Maximization of the Likelihood Function}
\label{sect:MaxLik}
To be able to maximize the likelihood over $\bm{\theta}$, it is useful to have no constraints and to ensure that $\bm{\theta}$ is unique. To do so, several reparameterizations are needed and we will represent that by the vector function $\bm{h}()$. First, to ensure the restrictions $0 \leq \phi_m \leq 1$ and $\sum_{m=1}^4 \phi_m = 1$, we define
\begin{equation*}
\label{eq:tau}
\bm{\phi} = \bm{h_\bm{\phi}(\tau)} = \left\{
\begin{array}{ll}
\exp(\tau_m)/(1 + \sum_{j = 1}^{3} \exp(\tau_j) ) & \mbox{for} \ m \in \{1, 2, 3\} ,\\
1 - \sum_{m=1}^{3} \exp(\tau_m)/(1 + \sum_{j = 1}^{3} \exp(\tau_j))  & \mbox{for} \ m = 4.
\end{array}
\right.
\end{equation*}
This reparametrization allows the constrained optimization over $\bm{\phi}$ to be replaced by the unconstrained optimization over $\bm{\tau}$. Secondly, a similar reparametrization is used for avoiding the sum one and nonnegativity constraints on the $S$ segment probabilities in $\bm{\pi}$ by $S - 1$ values through $\bm{\pi} = \bm{h_\pi(\sigma)}$.

Thirdly, there are sum zero constraints $\bm{C}_j\bm{\beta}_j = \bm{0}$ on the weights corresponding to the dummy coding of the categories belonging to each categorical variable $j$. Instead of using these constraints, one of the categories per categorical variable is assigned as a reference category and consequently that particular weight in $\bm{\beta}_j$ are set to zero, effectively excluding these weights from the optimization. This implies that the effect of the reference categories is included in the intercepts $\bm{\alpha}_u$. For notational convenience, it is useful to gather the intercepts in $\bm{\alpha}$ and weights $\bm{\beta}$ into a single vector $\bm{\gamma}$, that is, $\bm{\gamma}_u = [\bm{\alpha}'_u, \bm{\beta}'_u]'$ where $\bm{\beta}_u$ is the vector of weights with the values corresponding to reference categories fixed to zero. The transformation needed from the unique vector of parameters $\bm{\gamma}_u = [\bm{\alpha}'_u, \bm{\beta}'_u]'$ to the nonunique vector  $\bm{\gamma} = [\bm{\alpha}', \bm{\beta}']'$ is illustrated by the following example with $S = 3$, two categorical variables, and two numerical variables.  Then
\begin{eqnarray*}
   \bm{\gamma} = \left[
      		\begin{array}{c}
      			\bm{\alpha} \\
      			\bm{\beta}_1 \\
      			\bm{\beta}_2 \\
      			\beta_{3}\\
      			\beta_{4} 
      		\end{array}
      		\right]  
   = \bm{h}_{\bm{\gamma}}(\bm{\gamma}_u) = 
   \left[
   	\begin{array}{ccccc}
   		\bm{I}  & K_1^{-1}\bm{11}'& K_2^{-1}\bm{11}' & \bm{0} &\bm{0} \\
   		\bm{0} & \bm{I} - K_1^{-1}\bm{11}' & \bm{0} & \bm{0} & \bm{0}\\
   		\bm{0} & \bm{0} & \bm{I} - K_2^{-1}\bm{11}'  & \bm{0} & \bm{0}\\
   		\bm{0} & \bm{0} & \bm{0} & 1 & 0\\
   		\bm{0} & \bm{0} & \bm{0} & 0& 1\\
   	\end{array}
   	\right] \left[
   		\begin{array}{c}
   			\bm{\alpha}_u \\
   			\bm{\beta}_{u1} \\
   			\bm{\beta}_{u2} \\
   			\beta_{u3}\\
   			\beta_{u4} 
   		\end{array}
   		\right] = \bm{A\gamma}_u,
\end{eqnarray*}
where $K_j$ is the number of categories for categorical variable $j$ and the matrices $\bm{I}, \bm{0}$, and $\bm{11}'$ are understood to be adapted to the corresponding sizes depending on the relevant lengths of the vectors $\bm{\alpha}_u$ and $\bm{\beta}_{uj}$.


Let 
$\bm{\theta}_u '=  [\bm{\gamma}_u', \delta, \bm{\tau}', \bm{\sigma}']$ be the unconstrained vector of all uniquely defined parameters. Then, the reparametrization function $\bm{h}(\bm{\theta}_u) $ can be written as
\begin{eqnarray*}
	\bm{\theta} = \bm{h}(\bm{\theta}_u) = 
	[\bm{h}_{\bm{\gamma}}'(\bm{\gamma}_u), 
	\delta, 
	\bm{h}_{\bm{\phi}}'(\bm{\tau}), 
	\bm{h}_{\bm{\phi}}'(\bm{\sigma})]'. 
\end{eqnarray*}
Thus, the constrained maximization of $\bm{\theta}$ in (\ref{eq:ll}) is equivalent to the unconstrained maximization 
\begin{eqnarray*}
	L(\bm{\theta}) = L(\bm{h}(\bm{\theta}_u) = L_u(\bm{\theta}_u).
\end{eqnarray*}
To maximize $ L_u(\bm{\theta}_u)$ over $\bm{\theta}_u$, we use the BFGS quasi-Newton algorithm as implemented in the \texttt{optimx} package in \texttt{R}. After convergence, one step of the Newton-Raphson method with a numerically approximated Hessian is performed to ensure the gradient to be close to zero.

%
%

It is well known that for maximum likelihood estimation, at a maximum $\bm{\theta}_u^*$, the parameters are normally distributed $\bm{\theta}_u \sim N(\bm{\theta}_u^*, \bm{\Sigma}_u)$ where $\bm{\Sigma}_u = -(\nabla^2 L_u(\bm{\theta}_u^*)^{-1})$ is the inverse of the negative Hessian of $L_u()$ evaluated at $\bm{\theta}^*_u$. As we would like to have the covariance matrix $\bm{\Sigma}_{\bm{\theta}}$ of $\bm{\theta}^* = \bm{h}(\bm{\theta}_u^*)$, the Delta method is applied so that
\begin{eqnarray}
\bm{\theta}^* \sim N(\bm{h}(\bm{\theta}_u), \bm{\Sigma}_{\bm{\theta}} ) 
\label{for:delta_distr}
\end{eqnarray}
where  $\bm{\Sigma}_{\bm{\theta}} = \nabla \bm{h}'(\bm{\theta}_u)\bm{\Sigma}_u \nabla \bm{h}(\bm{\theta}_u)$ and $\nabla \bm{h}(\bm{\theta}_u)$ is the Jacobian matrix of $\bm{h}()$, that is, 
\begin{eqnarray*}
	\nabla \bm{h}(\bm{\theta}_u) = \left[
	\begin{array}{cccc}
		\nabla \bm{h}_{\bm{\gamma}}(\bm{\gamma}_f^*) & 0 & \bm{0} & \bm{0} \\
		\bm{0} & 1 & \bm{0} & \bm{0} \\
		\bm{0} & 0 & \nabla \bm{h}_{\bm{\phi}}(\bm{\tau}_u^*) & \bm{0} \\
		\bm{0} & 0 & \bm{0} & \nabla \bm{h}_{\bm{\pi}}(\bm{\sigma}_u^*) \\
	\end{array}
	\right], 
\end{eqnarray*}
where $\nabla \bm{h}_{\bm{\gamma}}(\bm{\gamma}_f^*) = \bm{A}$, $\nabla \bm{h}_{\bm{\phi}}(\bm{\tau}_u^*)$, and $\nabla \bm{h}_{\bm{\pi}}(\bm{\sigma}_u^*)$ are the Jacobians for $\bm{h}_{\bm{\gamma}}()$, $\bm{h}_{\bm{\phi}}()$, and $\bm{h}_{\bm{\pi}}()$, rerspectively. The standard errors of the model parameters in $\bm{\theta}$ are the square roots of the diagonal elements of the covariance matrix in (\ref{for:delta_distr}). 

\section{Significance Testing for Weighted Means}
\label{sect:sigtest}
Obtaining a test for the differences between weighted means as presented in Table~\ref{tab:segChar}, some nonstandard steps are needed. Let $\bm{P}$ be the $n \times S$ matrix of the a posteriori probabilities from the CMM. Then, regress the a posteriori probabilities in $\bm{P}$ on a CBCL symptom subscale without intercept through OLS. The obtained regression coefficients $\bm{\psi}$ can be transformed to the weighted averages as follows
\begin{equation*}
\bm{\psi}^* = \bm{B}\bm{\psi}, \text{ with } \bm{B} = \text{Diag}(\bm{P}'\bm{1})^{-1} \bm{P}'\bm{P},
\end{equation*}
where the operator Diag(.) transforms a vector into a diagonal matrix. We can do a Wald test with null hypothesis that the weighted means are the same ($\bm{\psi}^* = c\bm{1}$). The standard errors needed for the Wald test can be derived from the diagonal elements of the covariance matrix $\bm{\Sigma}_{\bm{\psi}^*} = \bm{B}'\bm{\Sigma_{\bm{\psi}}} \bm{B}$, where $\bm{\Sigma_{\bm{\psi}}}$ is the original covariance matrix obtained from the linear regression.

\section{CMM Expected Values}
\label{sec:pred}
As the CMM provides a probability distribution for each number of cards turned over on each trial, obtaining predictions from a Censored Mixture Model (CMM) is not straightforward. Therefore, we choose the expectation as a point estimate for the predicted value. These expectations can be obtained from the following steps. First, the estimated regression coefficients $\hat{\alpha}_s$ and $\hat{\beta}$ are used to compute the linear combination
\begin{equation*}
	\hat{\mu}_{its} = \log(\exp(\hat{\alpha}_s + \bm{x}'_{it}\bm{\hat{\beta}}) +1).
\end{equation*}
Second, for each individual, in each trial, and for each segment we can compute the probabilities of all possible outcomes $y \in \{0, ..., 32\}$,
\begin{equation}
\label{eq:probs}
	\Pr(Z_{it} = \ell \mid \hat{\mu}_{its}, \hat{\delta}) = \left\{
	\begin{array}{ll}
	\hat{\phi}_1 f(0 \mid \hat{\mu}_{its}, \hat{\delta}) + \hat{\phi_2}  & \text{if } \ell = 0,\\
	\hat{\phi}_1 f(\ell \mid \hat{\mu}_{its}, \hat{\delta}) + \frac{1}{|A|} \hat{\phi_3} & \text{if } \ell \in A,\\
	\hat{\phi}_1 f(31 \mid \hat{\mu}_{its}, \hat{\delta}) + \hat{\phi_4}  & \text{if } \ell = 31,\\
	\hat{\phi}_1 f(32 \mid \hat{\mu}_{its}, \hat{\delta}) & \text{if } \ell = 32,\\
	\hat{\phi}_1 f(\ell \mid \hat{\mu}_{its}, \hat{\delta}) & \text{for all other values } \ell,
	\end{array} 
	\right.
\end{equation}
where $f(\ell \mid \hat{\mu}_{its}, \hat{\delta})$ is the probability mass function of the negative binomial distribution, $A = \{4, 8, 10, 12, 16, 20, 24\}$, $|A|$ is the cardinality of set $A$, and $\hat{\phi}_j$ is the estimate of $\phi_j$. The distributions in Figures \ref{fig:uncen}, \ref{fig:cen}, \ref{fig:uncenOofS}, and \ref{fig:cenOofS} are obtained by weighting for the segments, using the estimated prior segment probability $\hat{\pi}_s$, and then summing over the individuals, trials, and segments
\begin{equation*}
	h_{it}(y) = \sum_{i = 1}^{N} \sum_{t=1}^{T} \sum_{s = 1}^{S} \hat{\pi}_s \Pr(Z_{it} = \ell \mid \hat{\mu}_{its}, \hat{\delta}) \ \ \ \ \forall \ \ell \in \{0, ..., 32\}.
\end{equation*}
Note that the right graphs of these figures are multiplied by the probability of being (un)censored to get the same scale as the left graph.

The expected values per person and per trial can be obtained from the probabilities in \eqref{eq:probs}, that is,
\begin{equation*}
	\hat{y}_{it} = \sum_{s = 1}^{S} \hat{\pi}_s \left(\sum_{\ell = 0}^{M} \ell \Pr(Z_{it} = \ell \mid \hat{\mu}_{its}, \hat{\delta}) + M\hat{\phi}_1F(M+1 \mid \hat{\mu}_{its}, \hat{\delta})\right).
\end{equation*}
The probability mass above $M$, denoted by $F(M+1 \mid \hat{\mu}_{its}, \hat{\delta})$, is added to the expected values as if it were the probability mass at $M$. Since the negative binomial distribution has an infinite upper bound, $M$ should be large to obtain an accurate expected value. The probability mass above $M = 100$ is smaller than 0.0001 and so will not have a meaningful effect on the expected value. Therefore, we choose $M = 100$.
\end{appendices}	

\end{document}